\documentclass[11pt]{article}
\usepackage{textcomp}
\usepackage{pdfsync}
\usepackage{fancyhdr}
\usepackage{amssymb}
\usepackage{amsmath}

\usepackage{graphicx}
\usepackage{latexsym}
\usepackage{appendix}
\usepackage{srcltx}
\usepackage{cite}

\newcommand{\rmd}{\mathrm{d}}

\begin{document}
\renewcommand{\thefootnote}{\fnsymbol{footnote}}

\vspace*{1.5cm}

\begin{center}
    {\LARGE \bf A constraint-free formulation of black hole \\[0.3cm]  thermodynamics from the field equations } \\[1.8cm]

    {\large Geonwoo Ahn,${}^{a}$~~Ijin Bae,${}^{b}$~~Geunyeong Jang,${}^{c}$~~and~~Yongjoon Kwon${}^{\ast}$} \\[0.6cm]
  
\footnotetext[1]{blackhole@sshs.hs.kr, \textsuperscript{a}23053@sshs.hs.kr, 
\textsuperscript{b}23044@sshs.hs.kr, 
\textsuperscript{c}23093@sshs.hs.kr}

    {\normalsize Physics Department, Seoul Science High School, Seoul 03066, Korea} \\[1.8cm]

   \vspace*{0.5cm}
    {\bf Abstract}
\end{center}
\footnotetext[2]{All authors contributed equally to this work.}
\footnotetext[3]{This paper has been accepted for publication in  the \textit{International Journal of Modern Physics D}.}

\begin{quote}
We develop a constraint-free formulation that generalizes Padmanabhan’s method for deriving the first law of black hole thermodynamics directly from the Einstein field equations. In previous studies, even for multi-horizon black holes, variations were restricted to the outer horizon by imposing an additional constraint, and the $P\,\mathrm{d}V$ term was introduced by multiplying the field equations evaluated at the outer horizon by the corresponding volume variation $\mathrm{d}V$.
However, since general variations of the black hole parameters shift both horizons, variations at both horizons must be taken into account. To this end, we propose multiplying the horizon field equations by the entropy variation $\mathrm{d}S$ under such unconstrained variations. We show that this method remains valid even in higher-derivative theories of gravity.
In addition, we find that $r_{\pm}$-based variation schemes generically break down for black holes characterized by three independent parameters $(M,J,Q)$. By working directly in the thermodynamic state space $(M,J,Q)$, we show that the Einstein field equations evaluated at the outer horizon can be also interpreted as the first law of black hole thermodynamics for general variations without imposing any additional constraints.
\end{quote}

%\noindent{\bf Keywords:} Hawking temperature; Einstein field equations; Black hole thermodynamics.

  \thispagestyle{empty}
\renewcommand{\thefootnote}{\arabic{footnote}}
\setcounter{footnote}{0}
\newpage

%%%%%%%%%%%%%%%%%%%%%%%%%%%%%%%%%%%%%%%%%%%%%%%%%%%%%%%%%%%%%%%%%%%%%%%
\section{Introduction}
%%%%%%%%%%%%%%%%%%%%%%%%%%%%%%%%%%%%%%%%%%%%%%%%%%%%%%%%%%%%%%%%%%%%%%%
The study of black holes has long focused on uncovering their profound connections to thermodynamic laws.
In particular, extensive research has investigated phenomena related to Hawking radiation and black hole thermodynamics.
In Ref.~\cite{Jacobson:1995ab}, it was shown that imposing the Clausius relation \(\delta Q = T\,\delta S\) on all local Rindler horizons—together with \(S\propto A\) and the Unruh temperature—yields the Einstein field equations, with the cosmological constant appearing as an integration constant.
Subsequently, Padmanabhan~\cite{Padmanabhan:2002sha} showed that the Einstein field equations, when evaluated at the black hole horizon, can be interpreted as the first law of black hole thermodynamics. 
He later refined this horizon thermodynamics, extended it to a broader class of spacetimes, and applied it to higher-curvature theories~\cite{Padmanabhan:2002ma, Padmanabhan:2003gd, Padmanabhan:2005zk, Paranjape:2006ca, Padmanabhan:2015zmr}.
Over the years, this connection has been investigated across diverse spacetimes and in generalized theories of gravity~\cite{Cai:2009ph, Wu:2009wp}.
 The essential idea underlying both the proposal of Ref.~\cite{Padmanabhan:2002sha} and its subsequent developments is that multiplying the \(G^{r}{}_{r}\) component of the field equations, evaluated at the horizon, by the infinitesimal change in the horizon volume \(\mathrm{d}V\) (or, equivalently, by the infinitesimal variation of the outer-horizon radius, \(\mathrm{d}r_{+}\), under an additional constraint that relates the two horizons) can be recast into the first-law form of black-hole thermodynamics.
This idea was first applied to the Schwarzschild black hole~\cite{Padmanabhan:2002sha}, and later extended to the Kerr–Newman~\cite{Kothawala:2007em} and rotating BTZ black holes~\cite{Akbar:2007qg}. 
However, prior studies  relating the field equations to thermodynamic laws for multi-horizon black holes imposed  specific constraints associated with the horizons~\cite{Kothawala:2007em,Akbar:2007qg,LarranagaRubio:2007uas,cadoni_cbtz,Akbar:2007zz,Larranaga:2008qw}.
For example,   the proposal was examined for the Kerr--Newman black hole by considering the particular variation \(\mathrm{d}J/J=\mathrm{d}M/M\)~\cite{Kothawala:2007em}.
In the analysis of the rotating BTZ black hole~\cite{Akbar:2007qg}, the angular velocity at the outer horizon is kept fixed during the variation, which implicitly imposes the particular relation between the two horizons, i.e., \(\mathrm{d}r_{-}=(r_{-}/r_{+})\,\mathrm{d}r_{+}\).
The same constraint was also adopted in their subsequent work on charged rotating BTZ black holes~\cite{Akbar:2007zz}. 
Also, for the same black holes, Padmanabhan’s proposal was confirmed by imposing a different constraint, $\rmd\Phi = 0$~\cite{Akbar:2007zz, LarranagaRubio:2007uas, cadoni_cbtz, Larranaga:2008qw}.

%%%%%%
However, we find that the approaches adopted in all previous studies have several limitations. First, they impose an additional constraint relating the two horizons, which effectively restricts variations to the outer horizon. 
Second, for non-spherically symmetric black holes such as Kerr and Kerr--Newman, multiplying the field equations at the outer horizon by the volume variation $\mathrm{d}V$  does not yield the first law of black hole thermodynamics under general variations at both horizons. This is because \(\mathrm{d}V\) is not proportional to the horizon-area variation \(\mathrm{d}A\), which in black-hole thermodynamics is replaced by the entropy variation \(\mathrm{d}S\).
Third, for black holes characterized by three independent thermodynamic variables $(M,J,Q)$, the thermodynamic state space is three-dimensional, whereas the horizon radii $r_{\pm}$ encode only two parameters. A formulation based exclusively on $r_{\pm}$ is therefore generically incomplete and cannot capture the fully general first law for such black holes.

This may explain why earlier studies~\cite{Kothawala:2007em,Akbar:2007qg,LarranagaRubio:2007uas,cadoni_cbtz,Akbar:2007zz,Larranaga:2008qw} of the Kerr--Newman and charged rotating BTZ black holes
resorted to imposing somewhat arbitrary constraints in order to establish the connection between the Einstein field equations and black hole thermodynamics, thereby circumventing the algebraic obstruction and suppressing the independent variation of $r_{-}$.

%%%%%%%%%%%%%%%%%%%%%%%%%%%

In this paper, we  develop a more general, constraint-free formulation that also applies to Kerr and Kerr--Newman black holes as the generalization of Padmanabhan's method. Also, we find that for black holes characterized by three independent parameters, the thermodynamic state space $(M,J,Q)$—rather than the horizon radii $r_{\pm}$—must be employed. Using our proposed formulation, we show that the first law of black hole thermodynamics can be derived directly from the field equations even for multi-horizon black holes in Einstein gravity and in higher-order gravity theories, without imposing any constraint.

Throughout this paper, we set \(c=\hbar=k_{\mathrm B}=1\).
In \(3{+}1\) dimensions we additionally take \(G=1\);
in \(2{+}1\) dimensions we instead adopt \(8G=1\).

 \vspace*{0.8cm}  
	%%%%%%%%%%%%%%%%%%%%%%%%%%%%%%%%%%%%%%%%%%%%%%%%%%%%%%%%%%%%
	\section{The Einstein Field Equations and the First Law of Black Hole Thermodynamics}
	%%%%%%%%%%%%%%%%%%%%%%%%%%%%%%%%%%%%%%%%%%%%%%%%%%%%%%%%%%%
In this section, we briefly review  the Padmanabhan's proposal for the Schwarzschild black hole~\cite{Padmanabhan:2002sha}, whose metric is 
\begin{eqnarray}
\rmd s^2 = -f(r) \,\rmd t^2 + \frac{1}{f(r)}\,\rmd r^2 + r^2\,(\rmd \theta^2 + \sin^2\theta \rmd \phi^2)\,,
\end{eqnarray}
where $f(r)=1-2M/r$. \\
For this metric, the relevant component of the Einstein field equations, ${G^{r}}_{r}$, takes the form of
\begin{eqnarray}
G^{r}{}_{r} = \frac{f'(r)}{r} + \frac{f(r)}{r^{2}} - \frac{1}{r^{2}} = 8\pi \, T^{r}{}_{r} \,.
\end{eqnarray}
By identifying ${T^{r}}_{r}$ with the radial pressure $P_{r}$ and multiplying both sides by $\rmd V/(8\pi)$, we obtain
\begin{eqnarray}
\Big(\frac{r f'(r)+f(r)-1}{2}\Big)\,\rmd r = P_{r}\,\rmd V  = P_{r}\,(4\pi r^{2})\,\rmd r \,.
\end{eqnarray}
At the horizon one identifies
\(T = f'(r_h)/(4\pi)\), \(S=\pi r_h^2\), and \(M=r_h/2\).
With these substitutions, the horizon field equation reduces to the
first-law  of black hole thermodynamics as follows:
\begin{eqnarray}
T\,\rmd S = \rmd M + P_r\,\rmd V \,.
\end{eqnarray}
For this metric, which is a vacuum solution, one obtains $T\,\rmd S = \rmd M$ since $P_r \equiv T^{r}{}_{r} = 0$.

To test the validity of the idea that the Einstein field equations evaluated at a horizon involve a thermodynamic interpretation equivalent to the first law of black-hole thermodynamics, numerous follow-up studies have been carried out~\cite{Padmanabhan:2002ma,Padmanabhan:2003gd,
Padmanabhan:2005zk,Paranjape:2006ca,Padmanabhan:2015zmr,Cai:2009ph,
Wu:2009wp,Kothawala:2007em,Akbar:2007qg,LarranagaRubio:2007uas,cadoni_cbtz,Akbar:2007zz,Larranaga:2008qw}.
However, for black holes with two horizons, these analyses typically do not consider general variations of both horizons. Instead, an additional constraint is assumed that effectively restricts the variation to the outer horizon only.

%%%%%%%%%%%%%%%%%%%%%%%%%%%%%%%%%%%%%%%%%%%%%%%%%%%%%%%%%%%%
	\section{Our proposal: A General Constraint-Free Derivation of the First Law for Multi-Horizon Black Holes }
	%%%%%%%%%%%%%%%%%%%%%%%%%%%%%%%%%%%%%%%%%%%%%%%%%%%%%%%%%%%%%%%%
Within Padmanabhan’s framework~\cite{Padmanabhan:2002sha}, a more natural and general procedure is to vary both the outer and inner horizons, $r_{+}$ and $r_{-}$, simultaneously when the black hole parameters are varied.
Even if the black hole volume is defined solely as a function of the outer-horizon radius, generic variations of $(M,J,Q)$ induce variations of both $r_{+}$ and $r_{-}$. Therefore, both horizons should participate in general variations. 

Moreover, for black holes such as the Kerr--Newman solution, which are characterized by three independent parameters \((M,J,Q)\), the thermodynamic variations are naturally expressed in terms of the differentials of these parameters. Since the two horizon radii satisfy \(r_{+}=r_{+}(M,J,Q)\) and \(r_{-}=r_{-}(M,J,Q)\), they do not form a complete set of independent variables. Consequently, for black holes with three independent parameters, a formulation based solely on variations of the horizon radii (\(\mathrm{d}r_{+}, \mathrm{d}r_{-}\)) is generally insufficient.

To address these issues, we propose a new formulation. By considering general variations without imposing any additional constraints, we show that the Einstein field equations can be recast into the thermodynamic first law.
Our method builds on the following key observation: the Hawking temperature is related to the surface gravity by
$T=\kappa/2\pi$, where $\kappa$ is determined by the metric of the black hole spacetime at the horizon.
In the static and spherically symmetric case, for instance, it is well known that the surface gravity is written as $\kappa=f'(r_h)/2$, so that $f'(r_h)$ directly determines the horizon temperature $T$.
Therefore, noting that an appropriate metric function at the horizon can be directly related to the Hawking temperature,  we propose multiplying the field equations by the entropy variation $\mathrm{d}S$, rather than by the volume variation $\mathrm{d}V$.
For example, in the ADM decomposition of the metric \cite{Arnowitt_2008,Gourgoulhon:2007ue},
\begin{equation}
\mathrm{d}s^2 = -N^2\,\mathrm{d}t^2
+ h_{ij}  \,\bigl(\mathrm{d}x^i+N^i  \, \mathrm{d}t\bigr) \, \bigl(\mathrm{d}x^j+N^j  \, \mathrm{d}t\bigr) \,,
\end{equation}
one finds that, for a static spacetime or for a stationary rotating black hole in a co-rotating frame, the horizon temperature $T$ is given by \cite{Padmanabhan:2003gd,Baines:2023cac} 
\begin{align}
T=\frac{\kappa}{2\pi}
= \frac{1}{4\pi}\left.\frac{\partial_r(N^2)}{\sqrt{N^2 h_{rr}}}\right|_{r_h}\,.
\end{align}
In this way, an appropriate horizon metric function can be identified with the horizon temperature. The combination $T\,\mathrm{d}S$ then arises naturally, making the connection between horizon geometry and thermodynamics explicit.

We now explain why, under general variations that shift both horizon radii, the field equations should be multiplied by $\mathrm{d}S$ rather than $\mathrm{d}V$. It should be emphasized that variations of the outer and inner horizons must be taken into account simultaneously. In Padmanabhan’s construction, multiplying the field equations by $\mathrm{d}V$ effectively amounts to multiplying by a geometric variation associated with the outer horizon, namely the outer-horizon area variation $\mathrm{d}A_{+}$. Or, to do so, an additional constraint was required.

General variations of the black hole parameters induce the displacements of both horizons. 
Treating the two horizon displacements on the same footing, as required by the virtual-displacement viewpoint, implies that a relevant geometric quantity $\mathfrak{S}$ should depend on both horizon radii, $\mathfrak{S}=\mathfrak{S}(r_{+},r_{-})$. 
More rigorously, since the horizon radii are generically determined by the three black hole parameters $(M,J,Q)$, $\mathfrak{S}$ may be regarded as a state function, $\mathfrak{S}=\mathfrak{S}(M,J,Q)$. 
This situation arises in particular for rotating black holes such as Kerr or Kerr--Newman, where general variations of $(r_{+},r_{-})$ or $(M,J,Q)$ shift both the outer and inner horizons. 
We therefore propose identifying the relevant geometric quantity $\mathfrak{S}$ with the entropy $S$, rather than with the volume $V$.
In addition, this choice remains valid even in higher-derivative gravity, where the black hole entropy is no longer proportional to the outer-horizon area. We elaborate on these points below.

In the Schwarzschild case, the infinitesimal volume variation $\mathrm{d}V$ is simply proportional to the infinitesimal horizon-area variation $\mathrm{d}A$:
\begin{align}
\mathrm{d}V = 4\pi r_{+}^{2}\,\mathrm{d}r_{+} = \frac{r_{+}}{2}\,\mathrm{d}A.
\end{align}
Hence, multiplying the field equations by $\mathrm{d}V$ is effectively equivalent to introducing the entropy variation, since $S=A/4$.
Likewise, for the Reissner-Nordström black hole, the horizon geometry is completely characterized by the areal radius \(r_{+}\). Because the spacetime is spherically symmetric, the associated spherical volume is then naturally fixed as a function of \(r_{+}\) alone, i.e., \(V = V(r_{+})\). Therefore, $\mathrm{d}V$ is simply proportional to $\mathrm{d}A$ as well.

For the Kerr black holes, however, the thermodynamic volume is not expressed only in terms of the outer horizon, since the spacetime is non-spherically symmetric and horizon geometry depends on both horizons.  
As a result, treating $\mathrm{d}V$ as a fundamental variation and multiplying the Einstein field equations by it is not well motivated. It is because the variation $\mathrm{d}V$ derived from $V=V(r_{+},r_{-})$ is, in general, not proportional to the variation $\mathrm{d}A$ derived from $A=A(r_{+},r_{-})$. This implies that multiplying the Einstein field equations by $\mathrm{d}V$ cannot, in general, produce the $T\,\mathrm{d}S$ term.
Specifically, $\mathrm{d}V$ follows directly from differentiating the expression for the  volume $V$ given in Ref.~\cite{Dolan:2011xt}:
 \begin{align}
dV=\Big(\frac{3r_+ + 2r_-}{6}\Big)\,dA-\frac{2\pi}{3}r_- \big(r_+ + r_-\big)\,dr_+
\end{align}
It can be checked that the infinitesimal variation $\mathrm{d}V$ is not proportional to the infinitesimal change in the horizon area $\mathrm{d}A$, and therefore neither is the entropy variation $\mathrm{d}S$.

  Also, for the Kerr-Newman black holes the outer-horizon area $A$ and the volume $V$ are  given by~\cite{Dolan:2011xt}  
\begin{align}
A &= 4\pi\left(r_+^{2}+a^{2}\right) 
=4\pi\Big(r_+(r_+ + r_-)-Q^{2}\Big)
\quad \textrm{and}\\
V
&=
\frac{2\pi}{3\,r_+}
\left[
\left(r_+^{2}+a^{2}\right)\left(2r_+^{2}+a^{2}\right)
+Q^{2}a^{2}
\right] 
=\frac{A}{6r_+}
\left(
r_+^{2}+\frac{A}{4\pi}
\right)
+\frac{2\pi Q^{2}}{3r_+}
\left(
\frac{A}{4\pi}-r_+^{2}
\right)\,,
\end{align}
where $r_\pm=M\pm\sqrt{\,M^{2}-a^2-Q^{2}\,}$ with $a\equiv {J}/{M}$. 
Therefore, it can be checked that   $\mathrm{d}V$ is also not proportional to $\mathrm{d}S$ as well as $\mathrm dA$.

Consequently, multiplying the horizon field equations by the volume variation $\mathrm{d}V$ does not, in general, yield the $T\,\mathrm{d}S$ term, and thus the field equations cannot be identified with the first law of black hole thermodynamics under general variations.
This helps to understand why, in Ref.~\cite{Kothawala:2007em}, the authors imposed the special condition $\mathrm{d}a=0$—rather than allowing fully general variations—to apply Padmanabhan’s method to the Kerr-Newman spacetime.

In addition, it is well known that neither the horizon area nor the horizon volume constitutes a fundamental thermodynamic quantity in black hole thermodynamics. Also, semiclassical quantization of black holes and broader attempts to quantize gravity indicate that entropy plays a more fundamental role than the horizon area~\cite{Kwon:2010km,Kwon:2011zza,Kwon:2011ey}. That is, once higher-curvature or quantum corrections are included~\cite{kwon2013quantization,Kwon:2010km,Jacobson:2005kr,Jacobson:1993xs}, area-based expressions are no longer appropriate. Accordingly, the entropy variation $\mathrm{d}S$ provides the appropriate and fully general thermodynamic quantity.
Motivated by these observations, we propose multiplying both sides of the field equations by $\mathrm{d}S$, which naturally incorporates the virtual displacements of both horizons on an equal footing.

Within this framework, we show below that the first law of black hole thermodynamics follows directly from the Einstein field equations for a variety of black-hole solutions in both Einstein gravity and higher-order gravity theories, without imposing any additional constraints.

	%%%%%%%%%%%%%%%%%%%%%%%%%%%%%%%%%%%%%%%%%%%%%%%%%%%%%%%%%%%%%%%
	\subsection{Kerr-Newman black hole}
	%%%%%%%%%%%%%%%%%%%%%%%%%%%%%%%%%%%%%%%%%%%%%%%%%%%%%%%%%%%%%%%%
The metric of the Kerr–Newman black hole is given by
\begin{eqnarray}
\rmd s^2 = -\frac{f(r)}{\rho^2}(\rmd t - a \sin^2\theta\,\rmd\phi)^2 
+ \frac{\rho^2}{f(r)}\,\rmd r^2 
+ \rho^2\,\rmd\theta^2 
+ \frac{\sin^2\theta}{\rho^2}\,(a\,\rmd t - (r^2 + a^2)\,\rmd\phi)^2 \,,\quad
\end{eqnarray}
where
\begin{eqnarray}
f(r) = r^2 - 2 M r + Q^2 + a^2 \quad \text{and} \quad
\rho^2 \equiv r^2 + a^2 \cos^2\theta\quad \text{with}\quad
a \equiv \frac{J}{M}\,.
\end{eqnarray}
The outer horizon radius is $r_h = M + \sqrt{M^2 - a^2 - Q^2}\,$.

The thermodynamic quantities associated with the outer horizon of the Kerr–Newman black hole are also well known:
\begin{eqnarray}
T = \frac{f'(r_h)}{4\pi(r_h^2 + a^2)} \qquad 
%A = 4\pi(r_h^2 + a^2)~, \qquad 
\text{and}\qquad
S = \frac{A}{4} = \pi(r_h^2 + a^2) \,.
\end{eqnarray}
For this metric, the $G^{r}{}_{r}$ component of the Einstein tensor and the corresponding component of the energy–momentum tensor at the horizon are given by
\begin{eqnarray}
G^{r}{}_{r} = \frac{a^2 + r_h f'(r_h) - r_h^2}{(r_h^2 + a^2 \cos^2\theta)^2} \qquad \text{and}\qquad
{T^{r}}_{r} = -\frac{Q^2}{8\pi (r_h^2 + a^2 \cos^2\theta)^2} \,.
\end{eqnarray}
The field equation $G^{r}{}_{r} = 8\pi {T^{r}}_{r}$ then leads to the following simple relation:
\begin{eqnarray}
\frac{r_h^2 - a^2 - Q^2}{r_h} - f'(r_h) = 0 \,.
\end{eqnarray}
Multiplying the both sides   by the entropy variation $\rmd S$ (and divided by $A$), 
\begin{eqnarray}
\left(\frac{r_h^2 - a^2 - Q^2}{r_h}\right)\frac{1}{4\pi(r_h^2 + a^2)}\,\rmd\!\left[\pi(r_h^2 + a^2)\right]
- \frac{f'(r_h)}{4\pi(r_h^2 + a^2)}\,\rmd\!\left[\pi(r_h^2 + a^2)\right] = 0 \,.
\end{eqnarray}
Using the standard identification \(T=\kappa/2\pi=f'(r_h)/\big(4\pi (r_h^2 + a^2)\big)\), this becomes 
\begin{eqnarray}
\left(\frac{r_h^2 - a^2 - Q^2}{r_h}\right)\frac{(r_h\,\rmd r_h + a\,\rmd a)}{2(r_h^2 + a^2) } -  {T\,\rmd S }  = 0 \,. \label{firstKN}
\end{eqnarray}
To express this equation in terms of $(M, J, Q)$, we need to rewrite $\rmd r_h$ and $\rmd a$ in terms of $\rmd M$, $\rmd J$, and $\rmd Q$. 
Then using the  relations 
\begin{eqnarray}
\rmd r_h = \frac{r_h}{r_h - M}\,\rmd M - \frac{a}{r_h - M}\,\rmd a - \frac{Q}{r_h - M}\,\rmd Q  \quad ~\text{and}~\quad \rmd J = a\,\rmd M + M\,\rmd a\,,
\end{eqnarray}
we obtain
\begin{eqnarray}
r_h\,\rmd r_h + a\,\rmd a = \frac{r_h^2 + a^2}{r_h - M}\,\rmd M - \frac{a}{r_h - M}\,\rmd J - \frac{Q r_h}{r_h - M}\,\rmd Q \,.
\end{eqnarray}
Therefore, using this, it can be easily checked that Eq.~(\ref{firstKN}) is just the first law of  thermodynamics  for the Kerr-Newman black holes
\begin{align}
T\,\rmd S 
&= \rmd M - \Omega\,\rmd J - \Phi\,\rmd Q \,,
\end{align}
where $\Omega = a / (r_h^2 + a^2)$ is the angular velocity at the outer horizon  and $\Phi= {Qr_h}/{(r_h^2+a^2)}$ is  the electric potential~\cite{poisson2004relativist, MTW1973}. 
Therefore, without imposing any constraints, we have derived the first law  from the field equations for black holes with three parameters $(M, J, Q)$.

%%%%%%%%%%%%%%%%%%%%%%%%%%%%%%%%	
	\subsection{Rotating BTZ black hole}
	%%%%%%%%%%%%%%%%%%%%%%%%%%%%%
As a black hole solution in (2+1)-dimensional gravity, the rotating BTZ spacetime corresponds to \(M,J\neq 0\) and \(Q=0\)~\cite{Akbar:2007qg}.
It therefore possesses two horizons.
In contrast to previous work~\cite{Akbar:2007qg}, we derive the first law for the rotating BTZ black hole without any assumption, following the approach outlined in the preceding section.

The metric of the rotating BTZ black hole is given by
\begin{eqnarray}
\rmd s^2 = -f(r)\,\rmd t^2 + \frac{\rmd r^2}{f(r)} + r^2\left(\rmd\phi - \frac{J}{2r^2}\,\rmd t\right)^2 \,,
\end{eqnarray}
where $f(r) = -M + \frac{r^2}{\ell^2} + \frac{J^2}{4r^2}$, and we adopt the BTZ gravitational units $c = 8G = 1$.\\[3pt] 
The thermodynamic quantities are given as
\begin{align}
    T = \frac{f'(r_h)}{4\pi}\,,\qquad S = 4\pi r_h \,,\qquad\text{and}\qquad \Omega = \frac{J}{2r_h^2}\,,
\end{align}
where the outer horizon radius is obtained as
\begin{eqnarray}
r_h^2 = \frac{1}{2}\left(\ell^2 M + \sqrt{\ell^4 M^2 - J^2 \ell^2}\right) \,.
\end{eqnarray}
The differential of the horizon radius is then given by
\begin{eqnarray}
2r_h\,\rmd r_h 
= \frac{1}{2}\left(\frac{2r_h^2 \ell^2\,\rmd M - \ell^2 J\,\rmd J}{2r_h^2 - \ell^2 M}\right). \label{BTZdr}
\end{eqnarray}
The     \((^r_{r})\)  component of the Einstein field equations is given by
\begin{eqnarray}
G^{r}{}_{r} - \frac{1}{\ell^2} = \pi {T^{r}}_{r} = 0 \,.
\end{eqnarray}
Evaluating the Einstein tensor at the outer horizon, one finds that this equation becomes
\begin{eqnarray}
\frac{J^2 + 2r_h^3 f'(r_h)}{4r_h^4} - \frac{1}{\ell^2} = 0 \,.  \label{rBTZEFeu}
\end{eqnarray}
Multiplying both sides by $\rmd S$, the above equation is rewritten as
\begin{eqnarray}
\frac{f'(r_h)}{4\pi}\,\rmd(4\pi r_h) = \left(-\frac{J^2}{4r_h^4} + \frac{1}{\ell^2}\right)(2r_h\,\rmd r_h) \,. \label{1stRBTZ}
\end{eqnarray}
%%%%%
%%%%%
Using Eq.~(\ref{BTZdr}), from Eq.~(\ref{rBTZEFeu}) we finally obtain the first law of black hole thermodynamics as follows:
\begin{eqnarray}
T\,\rmd S =\rmd M - \Omega\,\rmd J \,,
\end{eqnarray}
where $J = 2r_h^2 \Omega$ is used.

	%%%%%%%%%%%%%%%%%%%%%%%%%%%%%%%%%%%%%%%%%%%%%%%%%%%%%%%%%%%%%%%%
	\subsection{Charged Rotating BTZ black hole}
	%%%%%%%%%%%%%%%%%%%%%%%%%%%%%%%%%%%%%%%%%%%%%%%%%%%%%%%%%%%%%%%%
The charged rotating BTZ black hole is a general \((2+1)\)-dimensional black hole solution. 
Its metric has the same form as that of the rotating BTZ black hole, except for an additional logarithmic term in \(f(r)\) associated with the charge~\cite{Martinez:1999qi, Akbar:2007zz,LarranagaRubio:2007uas,cadoni_cbtz}:
\begin{eqnarray}
f(r) = -M + \frac{r^2}{\ell^2} + \frac{J^2}{4r^2} - \frac{\pi}{2}\,Q^2 \ln r \,.
\end{eqnarray}
Because of the logarithmic term in \(f(r)\), the horizon condition \(f(r)=0\) is transcendental and admits no closed-form solution. 
Consequently, the radii \(r_\pm\) cannot be expressed analytically as functions of \(M\), \(J\), and \(Q\).

As in the Kerr-Newman case, this limitation could be evaded by imposing auxiliary constraints in  previous works; for example, in Ref.~\cite{Akbar:2007zz} $\Phi$ and $\Omega$ were fixed. In practice, radius-based formulations that take \(r_\pm\) as fundamental variables generally do not permit a direct derivation of the first law from the field equations; they yield only differential relations linking \((\mathrm{d}M,\mathrm{d}J,\mathrm{d}Q)\) to \((\mathrm{d}r_+,\mathrm{d}r_-)\). We therefore reformulate the field equations directly in the state space \((M,J,Q)\), treating \(M,J,Q\) as independent variables and varying them accordingly; this is essential for a consistent, constraint-free derivation.

The thermodynamic quantities are given as
\begin{align}
    T = \frac{f'(r_h)}{4\pi},\quad S=4\pi r_h,\quad \Omega=J/2r_h^2 \quad \text{and} \quad \Phi = -\pi Q \ln r_h\,.
\end{align}
From the horizon condition $f(r_h)=0$, the black hole mass $M$ can be expressed as
\begin{eqnarray}
M =\frac{r_h^2}{\ell^2} + \frac{J^2}{4r_h^2} - \frac{\pi}{2} Q^2 \ln r_h  = \frac{r_h^2}{\ell^2} + \frac{1}{2}J\,\Omega + \frac{1}{2}\Phi \,Q \,,
\end{eqnarray}
where the angular velocity at the outer horizon is $\Omega =  {J}/{2r_h^2}$ and the electric potential is $\Phi = -\pi \,Q\, \ln r_h$.
Taking the differential of both sides yields
\begin{eqnarray}
\rmd M = \frac{2\,r_h\,\rmd r_h}{\ell^2} + \frac{J\,\rmd J}{2r_h^2} - \frac{J^2\,\rmd r_h}{2r_h^3} + \frac{1}{2}\rmd(\Phi \,Q) \,. \label{dM}
\end{eqnarray}
And, the  \((^r_{r})\)  component of the Einstein field equations, 
$G^{r}{}_{r} - \frac{1}{\ell^{2}} = \pi\,T^{r}{}_{r}$ with $T^{r}{}_{r} = - {Q^{2}}/{4r^{2}}$, 
reduces at the horizon $r = r_h$ to
\begin{eqnarray}
\frac{J^2 + 2\,r_h^3 \,f'(r_h)}{4\,r_h^4} - \frac{1}{\ell^2} = -\frac{\pi \,Q^2}{4\,r_h^2} \,.
\end{eqnarray}
Multiplying both sides by $\rmd S$, this equation becomes
\begin{eqnarray}
\frac{f'(r_h)}{4\pi}\,\rmd(4\pi r_h) 
= \left(\frac{2r_h}{\ell^2} - \frac{\pi Q^2}{2r_h} - \frac{J^2}{2r_h^3}\right)\rmd r_h \,.
\end{eqnarray}
Identifying the left-hand side with $T\,\rmd S$ and using Eq.~(\ref{dM}), we obtain the first law of black hole thermodynamics:
\begin{eqnarray}
T\,\rmd S = \rmd M - \frac{J\,\rmd J}{2r_h^2} - \frac{1}{2}\rmd(\Phi \, Q) - \frac{\pi Q^2}{2r_h}\,\rmd r_h \, = \rmd M \, - \Omega\,\rmd J - \Phi\,\rmd Q \,.
\end{eqnarray}

In the remaining sections, we apply our method to black holes in higher-derivative gravity and show that the field equations evaluated at the horizon take precisely the form of the first law of black hole thermodynamics.

 \vspace*{0.8cm}  
        %%%%%%%%%%%%%%%%%%%%%%%%%%%%%%%%%%%%%%%%%%%%%%%%%%%%%%%%%%%%%%%%
	\section{Black holes in New Massive Gravity}
	%%%%%%%%%%%%%%%%%%%%%%%%%%%%%%%%%%%%%%%%%%%%%%%%%%%%%%%%%%%%%%%%
New Massive Gravity (NMG) is  one of the higher-derivative gravity theories in (2+1) dimensions~\cite{Bergshoeff:2009hq}.
Within NMG, two well-known classes of black hole solutions are the rotating BTZ black holes and the new-type black holes.
As shown below, the NMG field equations evaluated at the horizon of these solutions can be recast into the first-law form of black hole thermodynamics.

In NMG, the action is given by~\cite{Bergshoeff:2009hq,Bergshoeff:2009aq}
\begin{eqnarray}
    \mathcal{S}_{\mathrm{NMG}}
    = \mathcal{S}_{\mathrm{EH}} + \mathcal{S}_{R^2} 
    = \frac{1}{16\pi G} \int d^{3}x \sqrt{-g}
       \Bigl[
           R + \frac{2}{\ell^{2}}
           + \frac{1}{m^{2}}
             \Bigl(
                 R_{\mu \nu} R^{\mu \nu}
                 - \frac{3}{8} R^{2}
             \Bigr)
       \Bigr].
\end{eqnarray}
The corresponding modified Einstein field equations are given by
\begin{eqnarray}
    G_{\mu \nu}
    - \frac{1}{\ell^{2}} g_{\mu \nu}
    + \frac{1}{2 m^{2}} K_{\mu \nu}
    = 0 \,, \label{NMGfeq}
\end{eqnarray}
where
\begin{align}
    K_{\mu \nu}
    = g_{\mu \nu}
      \Bigl(
          3 R_{\alpha \beta} R^{\alpha \beta}
          - \frac{13}{8} R^{2}
      \Bigr)
      + \frac{9}{2} R R_{\mu \nu}
      - 8 R_{\mu \alpha} R^{\alpha}{}_{\nu}
      + \frac{1}{2}
        \Bigl(
            4 \nabla^{2} R_{\mu \nu}
            - \nabla_{\mu} \nabla_{\nu} R
            - g_{\mu \nu} \nabla^{2} R
        \Bigr)\,.
\end{align}

                %%%%%%%%%%%%%%%%%%%%%%%%%%%%%%%%%%%%%%%%%%%%%%%%%%%%%%%%%%%%%%%%
	\subsection{Rotating BTZ black holes in NMG}
	%%%%%%%%%%%%%%%%%%%%%%%%%%%%%%%%%%%%%%%%%%%%%%%%%%%%%%%%%%%%%%%%
The metric of the rotating BTZ black hole is given by 
\begin{eqnarray}
    \rmd s^2 = -f(r)\, \rmd t^2 + \frac{\rmd r^2}{f(r)}
    + r^2 \left( \rmd \phi - \frac{J}{2r^{2}}\, \rmd t \right)^{2}\,,
    \label{RBTZmet}
\end{eqnarray}
where
\begin{eqnarray}
    f(r)
    = -M + \frac{r^{2}}{L^{2}} + \frac{J^{2}}{4r^{2}}
    = \frac{1}{r^{2}L^{2}}\,(r^{2} - r_{+}^{2})(r^{2} - r_{-}^{2})\,.
\end{eqnarray}
For this metric to satisfy the NMG field equations~(\ref{NMGfeq}), the effective AdS scale \(\ell\) must be related to the BTZ parameter \(L\) and the NMG mass parameter \(m\) as
\begin{eqnarray}
    \frac{1}{\ell^{2}}
    = \frac{1}{L^{2}}
      \left(
          1 - \frac{1}{4m^{2}L^{2}}
      \right).
    \label{lLrel}
\end{eqnarray}
And the \((^r_{r})\) component of the field equation is given by
\begin{eqnarray}
    {G^{r}}_{r} - \frac{1}{\ell^{2}} g^{r}{}_{r} + \frac{1}{2m^{2}} K^{r}{}_{r}= \frac{2f'(r)r^3 + J^2}{4r^4} - \frac{1}{L^{2}}\left(1 - \frac{1}{4m^{2}L^{2}}\right) + \frac{1}{2m^{2}}K^{r}{}_{r} =0 \,,  \qquad   \label{BTZeom}
\end{eqnarray}
where
\begin{align}
    K^{r}{}_{r} = \frac{1}{32 r^{8}}
    \Big[&\left( -8 f'''(r) r^8 - 8 f''(r) r^7 - 12 J^2 r^3 \right) f'(r) \nonumber \\ 
&+ \left( 16f(r)f'''(r)r^7+96J^2f(r)r^2+4 (f''(r))^2 r^8 - 40 J^2 f''(r) r^4 + 63 J^4\right) \Big]\,.
\end{align}
Note that the terms involving $f''(r)$ and $f'''(r)$ originate from higher-derivative corrections, which introduce additional contributions to the field equations.
\\
In the Einstein limit ($m\to\infty$), the thermodynamic variables of the BTZ black hole can be written entirely in terms of the horizon radii $r_{+}$ and $r_{-}$:
\begin{eqnarray}
    M = \frac{r_{+}^{2} + r_{-}^{2}}{L^{2}}, \quad
    J = \frac{2 r_{+} r_{-}}{L}, \quad
    \Omega = \frac{r_{-}}{r_{+} L}, \quad
    T = \frac{f'(r_{+})}{4\pi}, \quad \text{and}\quad
    S = 4 \pi r_{+} \,.
    \label{BTZvar}
\end{eqnarray}
First, let us consider Einstein limit case. At the outer horizon, using $f(r_\pm)=0$, Eq.~(\ref{BTZeom}) reduces to
\begin{align}
    &\frac{1}{64m^2}\left[\left( 4\left(f''(r_+)\right)^2-\frac{40J^2f''(r_+)}{r_+^4}+\frac{63J^4}{r_+^8}+\frac{16}{L^4} \right)- \left( 8f'''(r_+)+\frac{8f''(r_+)}{r_+}+\frac{12J^2}{r_+^5} \right)f'(r_+)\right]\nonumber \\ 
&+\left[\frac{2f'(r_+)r_+^3 + J^2}{4r_+^4} - \frac{1}{L^{2}}\right] =0  \,.
    \label{BTZeom2}
\end{align}
In Einstein gravity, only the second term remains. 
Hence Eq.~(\ref{BTZeom2}) in the Einstein limit simplifies to 
\begin{eqnarray}
\frac{2r_+^3 f'(r_+)+J^2}{4r_+^4} - \frac{1}{L^2} = 0 \,. \label{BTZeq1}
\end{eqnarray}
This is the identical structure of rotating BTZ black hole in Einstein gravity, where \(\ell\) is modified to \(L\). Using the same manner, it is straightforward to verify that Eq.~(\ref{BTZeq1}) can be rewritten as
\begin{eqnarray}
    T\, \mathrm{d}S -  \mathrm{d}M + \Omega\,  \mathrm{d}J = 0 \,.
\end{eqnarray}

Now, in NMG, the tensor $K_{\mu\nu}$ affects not only the equations of motion but also the conserved charges; 
hence, the thermodynamic quantities are given by the following ADT charges and the Bekenstein–Hawking–Wald entropy~\cite{Clement:2009gq,Nam:2010ub,Kwon:2011jz,Kim:2013qra}.
\begin{eqnarray}
    \mathcal{M} = \alpha M\,, \quad
    \mathcal{J} = \alpha J\,, \quad \text{and}\quad
    \mathcal{S} = \alpha S \,,
\end{eqnarray}
where $\alpha \equiv 1 +   ({1}/{2m^{2}L^{2}} )\,$.
\\
 Nevertheless, with the following expressions derived by using the condition $f(r_\pm)=0$\,,
\begin{align}
f'(r_+)
&= \frac{2r_+}{L^{2}} - \frac{J^{2}}{2 r_+^{3}}
 =  \frac{2\!\left(r_+^{2}-r_-^{2}\right)}{L^{2} r_+}\,, \quad
f''(r_+)
= \frac{2}{L^{2}} + \frac{3J^{2}}{2 r_+^{4}}
 =\frac{2}{L^{2}} + \frac{6 r_-^{2}}{L^{2} r_+^{2}}\,, \nonumber \\ 
\text{and}\qquad 
f'''(r_+)
&= -\frac{6J^{2}}{r_+^{5}}
= -\frac{24 r_-^{2}}{L^{2} r_+^{3}}\,,  
\end{align}
it can also be shown that the Einstein field equation at the horizon, Eq.~(\ref{BTZeom2}) simplifies to
\begin{eqnarray}
    \left( 1 + \frac{6ML^2-7r_+^2}{2m^2L^2r_+^2} \right)\left[ f'(r_+)-\frac{2\,(2r_+^2-ML^2)}{L^2r_+} \right]   \label{BTZeom3}=0\,.
\end{eqnarray}
Multiplying Eq.~(\ref{BTZeom3}) by $ \mathrm{d}\mathcal{S} = 4\pi\alpha\, \mathrm{d}r_{+}$ yields
\begin{eqnarray}
    \frac{f'(r_+)}{4\pi}  \mathrm{d}\mathcal{S}
    - \frac{1}{2\pi L^2}\frac{2r_+^2 - ML^2}{r_+}\left(4\pi\alpha  \,\mathrm{d}r_+\right)
    = 0\,, \label{BTZeq2}
\end{eqnarray}
so that the first term can be naturally identified with \(T\,\rmd \mathcal{S}\) 
in the first law of thermodynamics.
To express the result in terms of the black hole mass \(\mathcal{M}\) and angular momentum \(\mathcal{J}\),
one finds the relation from 
$r_+^2 = \tfrac{L}{2\alpha}\bigl(\mathcal{M}L+\sqrt{\mathcal{M}^2L^2-\mathcal{J}^2}\bigr)$:
\begin{eqnarray}
    r_+\,  \mathrm{d}r_+
    = \frac{L^2}{4\alpha} \Big( \frac{  2\alpha r_+^2\,  \mathrm{d}\mathcal{M} - \mathcal{J}\,  \mathrm{d}\mathcal{J} }{2\alpha r_+^2-\mathcal{M}L^2}\Big)
     \,.
\end{eqnarray}
Using this, Eq.~(\ref{BTZeq2}) takes precisely the form of the first law of black hole thermodynamics:
\begin{eqnarray}
    T\, \rmd \mathcal{S} - \rmd \mathcal{M} + \Omega\, \rmd \mathcal{J} = 0\,.
\end{eqnarray}
Thus, in NMG  higher derivative corrections are entirely absorbed by redefining the thermodynamic quantities,
\((M,J,S)\!\to\!(\mathcal{M},\mathcal{J},\mathcal{S})\).
With these redefinitions, the near horizon contributions
(e.g., the temperature gradient) are incorporated, and the first law
structure is preserved.

                    %%%%%%%%%%%%%%%%%%%%%%%%%%%%%%%%%%%%%%%%%%%%%%%%%%%%%%%%%%%%%%%%
	\subsection{New type black holes in NMG}
	%%%%%%%%%%%%%%%%%%%%%%%%%%%%%%%%%%%%%%%%%%%%%%%%%%%%%%%%%%%%%%%%
At the special point \( \ell^{2} = 1/m^{2} = 2L^2 \), 
the field equations (\ref{NMGfeq}) admit both the new-type black hole and the non-rotating BTZ black hole as solutions~\cite{Bergshoeff:2009hq, Bergshoeff:2009aq}.
We now focus on the new-type black hole. 
The analysis proceeds in complete analogy with the BTZ case in the previous section, 
and its metric is  
\begin{eqnarray}
    \rmd s^{2} = -f(r)\,\rmd t^{2} + \frac{\rmd r^{2}}{f(r)} + r^{2}\,\rmd \phi^{2}\,,  \label{metricNT}
\end{eqnarray}
where 
\begin{align}
f(r) = \frac{r^{2}}{L^{2}} + \frac{b\,r}{L} + c 
     = \frac{1}{L^{2}}(r - r_{+})(r - r_{-})\,.
\quad
\end{align}
In particular, when \(b = 0\) and \(c < 0\), the solution reduces to the non-rotating BTZ black hole. 
The thermodynamic quantities of this black hole, including the Bekenstein-Hawking-Wald entropy, can be expressed in terms of the horizon radii \(r_{\pm}\) as \cite{Oliva:2009ip,Kwon:2011ey,Nam:2010ub,Kwon:2011jz,Kim:2013qra}
\begin{eqnarray}
    M = \frac{(r_{+} - r_{-})^{2}}{2L^{2}}\,, 
    \qquad 
    S = 4\pi(r_{+} - r_{-})\,, 
    \qquad  \text{and}\qquad
    T = \frac{f'(r_{+})}{4\pi} \,.
    \label{NTMS}
\end{eqnarray}
For the metric (\ref{metricNT}), the \((^r_r)\) component of the field equation (\ref{NMGfeq}) is  
\begin{align}
    G^{r}{}_{r} - \frac{1}{2L^{2}}g^{r}{}_{r} + L^{2}\,K^{r}{}_{r}
    = \frac{f'(r)}{2r}-\frac{1}{2L^2}+L^2\, K^{r}{}_{r} =0\,,
    \label{NTEOM2}
\end{align}
where  
\begin{align}
K^{r}{}_{r} = \left[ -\frac{f'(r)\,f'''(r)}{4} + \frac{\big(f''(r)\big)^2}{8} + \frac{f(r)\,f'''(r)}{2r} - \frac{f'(r)\,f''(r)}{4r}\right] \,.\label{NTEOM2_1}
\end{align}
Note that \(f''(r_+)=2/L^2\) is constant because the metric function \(f(r)\) is quadratic. Thus, the first term in \(K^{r}{}_{r}\)  vanishes and, more interestingly, the last term of \(K^{r}{}_{r}\) cancels against the \(G^{r}{}_{r}\) term for generic \(r\). Consequently, the field equation (\ref{NTEOM2}) contains no \(f'(r)\) term by which the temperature can be identified.
 To address this, we read the temperature from the \(G^{r}{}_{r}\) component alone, namely \(T=f'(r_+)/(4\pi)\).
One could in principle read the temperature from the \( f'(r) \) that appears inside \( K^{r}{}_{r} \). However, this   cannot be done together with the temperature identification from \( G^{r}{}_{r} \), because treating both terms as representing the temperature makes the \( f'(r) \) contribution vanish, rendering the identification untenable.
   This choice is justified by two observations: (i) in the limit \(m\!\to\!\infty\), the theory reduces to Einstein gravity, where \(f'(r_+)/(4\pi)\) yields \(T=\kappa/2\pi\); (ii) the factor \(f'(r_+)\) in Eq.~(\ref{NTEOM2_1}) comes from the higher–curvature tensor \(K^{r}{}_{r}\) and never appears by itself, but only as the product \(f'(r_+)f''(r_+)\). Since this composite coefficient is not the horizon derivative that defines the surface gravity, it should not be interpreted as the temperature.
Therefore, from Eq.~(\ref{NTEOM2}) at the outer horizon, multiplying by \(\mathrm{d}S\), one finds
\begin{eqnarray}
    \frac{f'(r_{+})}{4\pi}\,\mathrm{d}S
    - \frac{L^{2}\,f'(r_{+})\,f''(r_{+})}{8\pi}
      \left(\frac{4\pi L^{2}\mathrm{d}M}{r_+ - r_-}\right)
    + \left[\frac{L^{2}\,\big(f''(r_{+})\big)^{2}}{8} - \frac{1}{2L^{2}}\right]
      \frac{r_{+}}{2\pi}\,\mathrm{d}S = 0\,,\qquad
    \label{NTEOM3}
\end{eqnarray}
where the following relation is used:
\begin{eqnarray}
    (r_{+} - r_{-})\,\frac{ \mathrm{d}S}{4\pi} = L^{2}\, \mathrm{d}M \,.
    \label{NTrel}
\end{eqnarray}
With \(f'(r_+)f''(r_+)=2(r_{+}-r_{-})/L^{4}\), we find that Eq.~(\ref{NTEOM3}) reproduces the first law of black hole thermodynamics,  
\begin{align}
    T\,\mathrm{d}S - \mathrm{d}M = 0 \,.
\end{align}

        %%%%%%%%%%%%%%%%%%%%%%%%%%%%%%%%%%%%%%%%%%%%%%%%%%%%%%%%%%%%%%%%
	\section{Black holes in Topologically Massive Gravity}
	%%%%%%%%%%%%%%%%%%%%%%%%%%%%%%%%%%%%%%%%%%%%%%%%%%%%%%%%%%%%%%%%
In this section, we also show that, for black holes arising in topologically massive gravity (TMG)—one of the higher-derivative gravity theories—the Einstein field equations at the horizon likewise take the form of the first law of thermodynamics under our proposed formulation.

The action of TMG consists of the Einstein–Hilbert term with a negative cosmological constant and a gravitational Chern–Simons term, and is given by~\cite{Deser:1981wh}
\begin{align}
\mathcal{S}_{\text{TMG}} 
&= \mathcal{S}_{\text{EH}} + \mathcal{S}_{\text{CS}} \nonumber\\
&= \frac{1}{16\pi G} \int d^3x\,\sqrt{-g}\left( R + \frac{2}{\ell^2} \right)
 - \frac{\ell}{96\pi G\nu} \int d^3x\,\sqrt{-g}\,\varepsilon^{\lambda\mu\nu}
 \Gamma^{r}_{\lambda\sigma} \left( \partial_\mu \Gamma^{\sigma}_{r\nu} 
 + \frac{2}{3}\Gamma^{\sigma}_{\mu\tau}\Gamma^{\tau}_{\nu r} \right)\,,
\end{align}
where $\nu$ is the coupling constant and $\varepsilon^{012} = +1/\sqrt{-g}$ is the Levi–Civita tensor. 
Varying this action with respect to the metric, the modified Einstein field equations are obtained as
\begin{align}
G_{\mu\nu} - \frac{1}{\ell^2} g_{\mu\nu} + \frac{\ell}{3\nu} C_{\mu\nu} = 0\,,   \label{CStermeom}
\end{align}
where $C_{\mu\nu}$ is the Cotton tensor defined as
\begin{align}
C_{\mu\nu} = \varepsilon_{\mu}{}^{\alpha\beta}\nabla_{\alpha}\left(
R_{\beta\nu} - \frac{1}{4} g_{\beta\nu} R \right)\,.
\end{align}
It is well known that the solutions to Eq.~(\ref{CStermeom}) include the BTZ black hole, which has a vanishing Cotton tensor, and the warped AdS black hole, for which the Cotton tensor is non-vanishing.

        %%%%%%%%%%%%%%%%%%%%%%%%%%%%%%%%%%%%%%%%%%%%%%%%%%%%%%%%%%%%%%%%
	\subsection{Rotating BTZ black holes in TMG}
	%%%%%%%%%%%%%%%%%%%%%%%%%%%%%%%%%%%%%%%%%%%%%%%%%%%%%%%%%%%%%%%%
The metric of the rotating BTZ black hole is given by 
\begin{eqnarray}   
    \rmd s^2 = -f(r)\, \rmd t^2 + \frac{\rmd r^2}{f(r)}
    + r^2 \left( \rmd \phi - \frac{J}{2r^{2}}\, \rmd t \right)^{2}\,,
\end{eqnarray}
where $f(r)=-M+\frac{r^2}{\ell^2}+\frac{J^2}{4r^2}\,$.
This black hole admits inner and outer horizons \(r_\pm\) defined by
\(f(r_\pm)=0\,\). Their radii, expressed in terms of \(M\) and \(J\), are
\begin{align}
r_\pm^2= {\frac{1}{2}\!\left(M\,\ell^{2}\pm\sqrt{M^{2}\,\ell^{4}-J^{2}\,\ell^{2}}\right)}\,,
\end{align}
From the horizon condition $f(r_+)=0$, the relation among the variations of $r_+$, $M$, and $J$ is obtained as
\begin{equation}
\mathrm{d}r_+=\frac{1}{r_+^2-\frac{J^2\,\ell^2}{4\,r_+^2}}
\left(\frac{1}{2}\,\ell^2r_+\,\mathrm{d}M-\frac{J\,\ell^2}{4\,r_+}\mathrm{d}J\right)\,.
\label{CsTmg}
\end{equation}
For the  rotating BTZ black hole, the Hawking temperature, entropy, and angular velocity are given by
\begin{equation}
T = \frac{f'(r_+)}{4\pi}\,, \qquad 
S = 4\pi r_+\,, \qquad \text{and}\qquad
\Omega = \frac{J}{2\,r_+^2}\,.
\end{equation}
In topologically massive gravity (TMG), the conserved charges receive
corrections from the gravitational Chern–Simons term. Accordingly, the ADT charges and Wald entropy with Chern-Simons correction are given by
\begin{eqnarray}
\mathcal{M}=M+\frac{J}{3\nu\ell}\,, \qquad 
\mathcal{J}=J+\frac{M\,\ell}{3\nu}\,, \qquad \text{and}\qquad
\mathcal{S}=\left(1+\frac{J\,\ell}{6\nu r_{+}^2}\right) S \,.
\end{eqnarray}
%%%%%
%%%%%
From the \((^r_{r})\)  component of the field equation~(\ref{CStermeom}),  
\begin{eqnarray}
G^r{}_r - \frac{1}{\ell^2} g^r{}_r + \frac{\ell}{3\nu} C^r{}_r = 0 \,,
\end{eqnarray}
we find that  at the outer horizon
\begin{eqnarray}
2 f'(r_+) r_+^3 + J^2 - \frac{4 r_+^4}{\ell^2} = 0 \,.
\end{eqnarray}
With the identification \(T=f'(r_+)/4\pi\), multiplying by \(\rmd S\) yields
\begin{align}
T\,\mathrm{d}\mathcal{S} &=\left(-\frac{J^2}{8\pi r_+^3}+\frac{r_+}{2\pi\ell^2} \right)
\left[4\pi\,\mathrm{d}r_+\,\left(1-\frac{J\ell}{6\nu r_+^2}\right)
      +4\pi\,\frac{\ell}{6\nu r_+}\,\mathrm{d}J\right]\\
&= \mathrm{d}M - \Omega\,\mathrm{d}J
         - \frac{\ell}{3\nu}\,\Omega\,\mathrm{d}M
         + \frac{1}{3\nu\ell}\,\mathrm{d}J \,,
\label{TMGadt}
\end{align}
where Eq.~(\ref{CsTmg}) and $\Omega =  {J}/{2r_+^2}$ are used. 
Expressed in terms of the ADT charges \(\mathcal{M}\) and \(\mathcal{J}\), this reads
\begin{eqnarray}
\begin{aligned}
T\,\mathrm{d}\mathcal{S}
%&= \mathrm{d}\!\left(M + \frac{J}{3\nu\ell}\right)   - \Omega\,\mathrm{d}\!\left(J + \frac{\ell M}{3\nu}\right)
    = \mathrm{d}\mathcal{M} - \Omega\,\mathrm{d}\mathcal{J}\,.
\end{aligned}
\end{eqnarray}
Finally, it is also shown that  the field equations can be recast into the form of the first law of black hole thermodynamics.
%

        %%%%%%%%%%%%%%%%%%%%%%%%%%%%%%%%%%%%%%%%%%%%%%%%%%%%%%%%%%%%%%%%
	\subsection{Warped AdS Black Holes  in TMG}
	%%%%%%%%%%%%%%%%%%%%%%%%%%%%%%%%%%%%%%%%%%%%%%%%%%%%%%%%%%%%%%%%
We now turn to another black–hole solution of TMG: the warped AdS black hole,
whose Cotton tensor is nonvanishing. Its metric is given  by \cite{Anninos:2008fx,Bouchareb:2007yx}   
\begin{equation}
\begin{aligned}
\frac{\rmd s^{2}}{\ell^{2}}
= \rmd t^{2}
  + \frac{\ell^{2}}{4\,R(r)^{2}\,N(r)^{2}}\,\rmd r^{2}
  + 2\,R(r)^{2} N_{\theta}(r)\, \rmd t\, \rmd \theta
  + R(r)^{2}\, \rmd \theta^{2} \,,
\end{aligned}
\end{equation}
where 
\begin{align}
R(r)^{2} &\equiv \frac{r}{4}\left[3\left(\nu^{2}-1\right) r+\left(\nu^{2}+3\right)\left(r_{+}+r_{-}\right)-4 \nu \sqrt{r_{+} r_{-}\left(\nu^{2}+3\right)}\right]\,,  \nonumber
\\
N(r)^{2} &\equiv \frac{\ell^{2}\left(\nu^{2}+3\right)\left(r-r_{+}\right)\left(r-r_{-}\right)}{4 R(r)^{2}}\,, 
\qquad
N_{\theta}(r) \equiv \frac{2 \nu r-\sqrt{r_{+} r_{-}\left(\nu^{2}+3\right)}}{2 R(r)^{2}} \,.\nonumber
\end{align}
The \((^r_{r})\)  component of the field equation~(\ref{CStermeom}) is obtained as
\begin{align}
G^{\,r}{}_{r}&-\frac{1}{\ell^{2}}\,g^{\,r}{}_{r}+\frac{\ell}{3\nu}\,C^{\,r}{}_{\,r}
=
\frac{1}{\ell^{4}\nu}\Bigg[
    \nu\Bigl(
        R(r)^{4}\bigl[N_{\theta}'(r)\bigr]^{2}\ell^{2}
        +4N(r)R(r)\,N'(r)R'(r)
        -\ell^{2}
    \Bigr) \nonumber
\\[2pt]
&+\frac{4}{3}\,R(r)^{2}\Bigl(
        -R(r)R'(r)\,N(r)^{2}
        +R(r)^{2}N(r)N'(r)
    \Bigr) N_{\theta}''(r)
\nonumber \\[2pt]
&+\frac{4}{3}\,N_{\theta}'(r)R(r)^2\Bigl(
        R(r)^{4}\bigl[N_{\theta}'(r)\bigr]^{2}\ell^{2}
        -3N(r)^{2}\bigl[R'(r)\bigr]^{2}
        +4N(r)R(r)\,N'(r)R'(r)
\nonumber \\[2pt]
&\hspace{35pt}-N(r)R(r)^{2}N''(r)
        +N(r)^{2}R(r)R''(r)
        -R(r)^{2}\bigl[N'(r)\bigr]^{2}
    \Bigr)
\Bigg]  =0 \,.
\end{align}
At the outer horizon, using the condition $N(r_+)^2=0\,$, this equation reduces to
\begin{equation}\label{warpedEom}
\begin{aligned}
0
&= \frac{1}{\ell^{4}\nu}\Biggl\{
\nu\Bigl(
  \ell^{2}R(r_{+})^{4}[N'_{\theta}(r_{+})]^{2}
  + 4\,N(r_{+})R(r_{+})N'(r_{+})R'(r_{+})
  - \ell^{2}
\Bigr)
\\
&\qquad\qquad
+ \tfrac{4}{3}\,N'_{\theta}(r_{+})\,R(r_{+})^{2}\Bigl(
    \ell^{2}R(r_{+})^{4}[N'_{\theta}(r_{+})]^{2}    + 4\,N(r_{+})R(r_{+})N'(r_{+})R'(r_{+})
  \Bigr)
\\
& \qquad\qquad
+ \tfrac{4}{3}\,R(r_{+})^{4}N(r_{+})N'(r_{+})\,N''_{\theta}(r_{+}) 
- \tfrac{4}{3}\,N'_{\theta}(r_{+})\,R(r_{+})^{4}\Bigl(
    N(r_{+})N''(r_{+}) + [N'(r_{+})]^{2}
  \Bigr)
\Biggr\}\,.
\end{aligned}
\end{equation}
%%%%%
%%%%%
%
The thermodynamic quantities of the warped $AdS_3$ black hole, such as the ADT charges and the Wald entropy with the Chern–Simons correction, can then be expressed in terms of the horizon radii  \(r_{\pm}\) as \cite{Anninos:2008fx}
\begin{align}
\mathcal{S} &= \frac{\pi \ell}{3\nu}\left[\left(9\nu^{2}+3\right)r_{+}-\left(\nu^{2}+3\right)r_{-}
-4\nu\sqrt{\left(\nu^{2}+3\right)r_{+}r_{-}}\right]\,,  \label{TMGwarpeden}
\\[4pt]
\mathcal{M} &= \frac{\left(\nu^{2}+3\right)}{3}
\left(r_{+}+r_{-}-\frac{1}{\nu}\sqrt{r_{+}r_{-}\left(\nu^{2}+3\right)}\right)\,,
\\[4pt]
\mathcal{J} &= \frac{\nu \ell \left(\nu^{2}+3\right)}{12}
\left[
\left(r_{+}+r_{-}-\frac{1}{\nu}\sqrt{r_{+}r_{-}\left(\nu^{2}+3\right)}\right)^{2}
-\frac{\left(5\nu^{2}+3\right)}{4\nu^{2}}\left(r_{+}-r_{-}\right)^{2}
\right]\,,
\\[4pt]
T &\equiv \frac{1}{2\pi\ell}\sqrt{g^{rr}}\,
\partial_{r}N\Big|_{r=r_+}
=\frac{R(r_+)N(r_+)N'(r_+)}{2\pi\ell^3}
=\frac{\left(\nu^{2}+3\right)}{4\pi\ell}
\frac{r_{+}-r_{-}}{2\nu r_{+}-\sqrt{\left(\nu^{2}+3\right)r_{+}r_{-}}}\,,    \label{TMGwarpedte}
\\[4pt]
\Omega &\equiv \frac{N_{\theta}(r_+)}{\ell}
=\frac{2}{2\nu r_{+}-\sqrt{\left(\nu^{2}+3\right)r_{+}r_{-}}}\,.  \label{TMGwarpedan}
\end{align}
Using the relation $\,T = {N(r_+)N'(r_+)R(r_+)}/{(2\pi\ell^{3})}\,$, 
Eq.~(\ref{warpedEom}) simplifies to
\begin{align}
\Big(
 & \tfrac{20}{3}\,N'_{\theta}(r_{+})\,R(r_{+})^{2}\,R'(r_{+})
  + 4\nu\,R'(r_{+})
  + \tfrac{4}{3}\,N''_{\theta}(r_{+})\,R(r_{+})^{3}
\Big) T \nonumber
\\[2pt]
&= -\frac{1}{6\pi\ell}\Big(
     4\,R(r_{+})^{6}\,[N'_{\theta}(r_{+})]^{3}
     + 3\nu\,R(r_{+})^{4}\,[N'_{\theta}(r_{+})]^{2}-3\nu
   \Big)
 + \tfrac{4}{3}\,N'_{\theta}(r_{+})\,R(r_{+})^3\,
   \left.\frac{\mathrm d T}{\mathrm d r}\right|_{r=r_{+}}\, .
\label{tmgwarpedEOM}
\end{align}
%%%%%
%
%
Multiplying both sides of Eq.~(\ref{tmgwarpedEOM}) by $\mathrm{d}\mathcal{S}$  and expressing    the thermodynamic variables (\(T,\,S,\,\Omega\)) and the metric functions
($N_\theta\,$, $R$) in terms of the ADT charges \((\mathcal{M},\mathcal{J})\) (see the Appendix), we find that Eq.~(\ref{tmgwarpedEOM}) reduces to
%%%%%
\begin{align}
    T\,\mathrm{d}\mathcal{S}
    &= \,\mathrm{d}\mathcal{M}
    - \frac{2}{\ell\nu
    \left[\left(\frac{3 \mathcal{M}}{\nu^{2}+3}\right)
    + \sqrt{\frac{4 \nu^{2}}{5 \nu^{2}+3}
    \left(\left(\frac{3 \mathcal{M}}{\nu^{2}+3}\right)^{2}
    -\frac{12 \mathcal{J}}{\nu\ell(\nu^{2}+3)}\right)}\right]}\,
    \mathrm{d}\mathcal{J} \nonumber\\
    &= \,\mathrm{d}\mathcal{M} - \Omega\,\mathrm{d}\mathcal{J}\,.
\end{align}
Thus, for the warped AdS black hole the field equations are equivalent to the standard thermodynamic form as well.

  \vspace*{0.8cm}  
	%%%%%%%%%%%%%%%%%%%%%%%%%%%%%%%%%%%%%%%%%%%%%%%%%%%%%%%%%%%%%%%%
	\section{Conclusion and Discussion}
	%%%%%%%%%%%%%%%%%%%%%%%%%%%%%%%%%%%%%%%%%%%%%%%%%%%%%%%%%%%%%%%%
Following Padmanabhan’s idea~\cite{Padmanabhan:2002sha}, many studies have investigated how the particular component of the field equations at the horizon encode the first law of black hole thermodynamics~\cite{Padmanabhan:2002ma,Padmanabhan:2003gd,Padmanabhan:2005zk,Paranjape:2006ca,Padmanabhan:2015zmr,
Cai:2009ph,Wu:2009wp,Kothawala:2007em,Akbar:2007qg,LarranagaRubio:2007uas,cadoni_cbtz,Akbar:2007zz,Larranaga:2008qw}. 
Even for black holes with two horizons, however, variations have typically been restricted to the outer horizon by imposing a constraint relating \(r_{+}\) and \(r_{-}\), so the inner horizon is not varied independently.
As a result, a constraint-free derivation accommodating general variations is still missing for establishing the equivalence between the horizon field equations and the first law of black hole thermodynamics.

In this work, we generalize Padmanabhan’s virtual-displacement prescription, in which the horizon field equations are multiplied by the outer-horizon volume variation $\mathrm{d}V$. For spherically symmetric black holes with a single event horizon, the virtual displacement of  the single horizon is fully encoded in the variation of the  horizon volume. Therefore, only in this case multiplying the field equations by the infinitesimal volume variation $\mathrm{d}V$ --- or, equivalently, by the corresponding area variation $\mathrm{d}A$ --- is well motivated.

Beyond the spherically symmetric cases, however, this procedure is no longer adequate.
In particular, for multi-horizon black holes, general variations of the black hole parameters
induce variations of both the outer and inner horizons, and these variations
should therefore be treated simultaneously, without imposing any additional
constraints.
%%%%%%
This is a natural consequence of the fact that when a particle carrying angular momentum or electric charge falls into the black hole, both the outer and inner horizons are simultaneously affected. A variation of the outer horizon alone is possible only when the black hole parameters are not independent but are instead related by a specific constraint equation. In particular, for the Kerr or Kerr–Newman black hole, unlike the spherically symmetric case, the entropy of the outer horizon depends on both horizons. This implies that a variation of the inner horizon alone can affect the entropy, even in the absence of any change in the outer horizon. Accordingly, general variations of both horizons must be taken into account.
%%%%%%%%%%%%%%%%%
%
In this consideration, however, $\mathrm{d}V$ is not generally proportional to
$\mathrm{d}A$ (or equivalently $\mathrm{d}S$ ). 
For example, for non-spherically symmetric Kerr black holes with two horizons, the thermodynamic volume depends not only on the outer horizon but also on the inner horizon, i.e., $V=V(r_+,r_-)$. Moreover, for Kerr--Newman black holes with two horizons  characterized by three independent parameters, the volume is given by $V = V(M, J, Q)$.

As a result, the infinitesimal volume variation $\mathrm{d}V$ is not proportional to the
horizon-area variation $\mathrm{d}A$, and multiplying the Einstein field equations by
$\mathrm{d}V$ therefore fails to yield the $T\,\mathrm{d}S$ term.
Consequently, under general variations at both horizons,
the first law of black hole thermodynamics cannot be derived from the horizon
field equations using the $\mathrm{d}V$-based prescription.

To overcome these limitations, we propose replacing the outer-horizon volume variation
$\mathrm{d}V$ with the entropy variation $\mathrm{d}S$. 
Since the entropy can, in general, be regarded as a function of both horizon radii, $S=S(r_{+},r_{-})$, its variation $\mathrm{d}S$ naturally incorporates the infinitesimal displacements of both horizons and thus allows for fully general variations without imposing any additional constraints.
Using the relation $T=\kappa/2\pi$ (with $\kappa=f'(r_h)/2$ for static metrics), multiplying the horizon field equations by $\mathrm{d}S$ directly produces the $T\,\mathrm{d}S$ term, while the remaining contributions combine to yield $\mathrm{d}M-\Omega\,\mathrm{d}J-\Phi\,\mathrm{d}Q$. In this way,  we show that the Einstein field equations evaluated at the horizon reproduce the first law of black hole thermodynamics in a constraint-free manner.

In particular, for black holes characterized by three independent thermodynamic variables $(M,J,Q)$, we find that a description based solely on the horizon radii $r_{+}$ and $r_{-}$ is no longer sufficient. Instead, variations should be formulated directly in the thermodynamic state space, $\mathrm{d}S=\mathrm{d}S(M,J,Q)$. We show that, for both Kerr--Newman and charged BTZ black holes, the thermodynamic first law follows from our procedure under fully general variations, without imposing any additional constraints.

We further verify that our construction is not restricted to Einstein gravity, but extends to three-dimensional higher-derivative gravity theories, including new massive gravity (NMG) and topologically massive gravity (TMG). 
In such higher-order gravity theories, the entropy is generally not proportional to the outer-horizon
area, but depends on both horizons. We show that our proposal still holds for various black holes
even in these extended gravity theories under general variations of both horizons.

%%%%%%%%
In this paper, we show that multiplying the field equation at the horizon by $\mathrm{d}S$, rather than $\mathrm{d}V$, reproduces the first law of black hole thermodynamics.  In Padmanabhan's approach, the ${T^{r}}_{r}$ component of the field equation is interpreted as a radial pressure $P$, and multiplying the field equation by $\mathrm{d}V$  yields  the $P\,\mathrm{d}V$ term that appears in the ordinary thermodynamic identity. However, the presence of a nonvanishing ${T^{r}}_{r}$ reflects the contribution of matter fields --- such as the electromagnetic field --- that ultimately give rise to terms such as $\Phi\,\mathrm{d}Q$ in black hole thermodynamics. This does not imply that the  $P\,\mathrm{d}V$ term  is physically equivalent to $\Phi\,\mathrm{d}Q$. Indeed, in the black hole first law, the work terms correspond to $\Omega\,\mathrm{d}J + \Phi\,\mathrm{d}Q$, which  play the role of the $P\,\mathrm{d}V$ term in  ordinary thermodynamics. Also, the term $\Omega\, \mathrm{d}J$ does not arise from  the $P \,\mathrm{d}V$ term.  More importantly, multiplying by $\mathrm{d}V$ does not generically lead to the laws of black hole thermodynamics.

%%%%%%%%%%
On the other hand, the notion of thermodynamic volume arises  rather naturally within the framework of extended phase space thermodynamics, because  the black hole mass $M$ is interpreted not as the internal energy $E$, but as the enthalpy $H= E + P\,V$, where  the pressure $P = -\Lambda/(8\pi)$, related to  the cosmological constant $\Lambda$, is treated as a thermodynamic variable.  Since the first law in extended phase space thermodynamics is given by $\mathrm{d}H = T\,\mathrm{d}S + \Omega\,\mathrm{d}J + \Phi\,\mathrm{d}Q + V\,\mathrm{d}P$ \cite{Dolan:2011xt, Kastor:2009wy, Dolan:2012jh}, by identifying $M = H$ and comparing with the total differential of $M=M(S, J, Q, P)$, one finds that the thermodynamic volume is given by $V = \left(\frac{\partial M}{\partial P}\right)_{S,J,Q}$. 
When all black hole parameters including the cosmological constant are allowed to vary,  the entropy may be regarded as a function $S=S(M, J, Q, P)$ in extended phase space thermodynamics. Even in this setting, our formalism would reproduce the extended first law of black hole thermodynamics, $ T\,\mathrm{d}S= \mathrm{d}M - \Omega\, \mathrm{d}J -\Phi\, \mathrm{d}Q - V\, \mathrm{d}P$.
%%%%%%%%%%%%%%%%%%%%%

Our results reinforce the gravity–thermodynamics correspondence by considering more general variations. Our formalism implies that this correspondence is not tied to any particular component of the field equations evaluated at the horizon.
Rather, whenever a horizon-local geometric quantity determines the temperature, an equation that isolates this quantity can encode thermodynamic information, based on the identification $T=\kappa/2\pi$ linking thermodynamics to horizon geometry.

%{\acknowledgment}
%%%%%%%%%%%%%%%%%%%%%%%%%%%%%%%%%%%%%%%%%%%%%%%%%%%%%%%%%%%%%%%%%
\section*{Acknowledgements}
%%%%%%%%%%%%%%%%%%%%%%%%%%%%%%%%%%%%%%%%%%%%%%%%%%%%%%%%%%%%%%%%%

%\ack

This work was supported by the Research $\&$ Education program of Seoul Science High School(SSHS) grant funded by the Seoul Metropolitan Office of Education(SMOE) (No.~2024-RnE-physics5). YK thanks Chaewon Jeong in Physics Department in Seoul National University for   the valuable comments and assistance provided during the early stages of this project.

\newpage
\appendix
%%%%%%%%%%%%%%%%%%%%%%%%%%%%%%%%%%%%%%%%%%%%%%%%%%%%%%%%%%%%%%%%%%%%%
 \renewcommand{\theequation}{A.\arabic{equation}}
  \setcounter{equation}{0}
\section{Field equation for warped AdS black holes in TMG}
%%%%%%%%%%%%%%%%%%%%%%%%%%%%%%%%%%%%%%%%%%%%%%%%%%%%%%%%%%%%%%%%%%%%%%
The metric of the warped AdS black hole is given by \cite{Anninos:2008fx,Bouchareb:2007yx}    
\begin{equation}
\begin{aligned}
\frac{ds^{2}}{\ell^{2}}
= dt^{2}
  + \frac{\ell^{2}}{4\,R(r)^{2}\,N(r)^{2}}\,dr^{2}
  + 2\,R(r)^{2} N_{\theta}(r)\, dt\, d\theta
  + R(r)^{2}\, d\theta^{2} \,,
\end{aligned}
\end{equation}
where 
\begin{align}
R(r)^{2} &\equiv \frac{r}{4}\left[3\left(\nu^{2}-1\right) r+\left(\nu^{2}+3\right)\left(r_{+}+r_{-}\right)-4 \nu \sqrt{r_{+} r_{-}\left(\nu^{2}+3\right)}\right]\,, \nonumber
\\
N(r)^{2} &\equiv \frac{\ell^{2}\left(\nu^{2}+3\right)\left(r-r_{+}\right)\left(r-r_{-}\right)}{4 R(r)^{2}}\,, 
\qquad
N_{\theta}(r) \equiv \frac{2 \nu r-\sqrt{r_{+} r_{-}\left(\nu^{2}+3\right)}}{2 R(r)^{2}} \,.\nonumber
\end{align}
Each of the horizon radii \(r_+\) and \(r_-\) can be expressed in terms of the ADM mass \(\mathcal{M}\) and angular momentum \(\mathcal{J}\), as follows:
\begin{align}
r_{\pm} &= \frac{1}{2}\Big( X \;\pm\; (C-A) \Big)\,.
\end{align}
where
\begin{align}
A &\equiv \frac{3\mathcal M}{\nu^{2}+3},
\quad
B \equiv \frac{12\,\mathcal J}{\nu\,\ell\,(\nu^{2}+3)},
\quad C  \equiv A
+ \frac{2\nu}{\sqrt{5\nu^{2}+3}}\sqrt{A^{2}-B} \,, \nonumber
\\ 
\text{and} \qquad 
X& \equiv
\frac{4\nu^2A}{3(\nu^{2}-1)}
\;+\;
\frac{2\nu\,\sqrt{\nu^{2}+3}}{3(\nu^{2}-1)\sqrt{5\nu^2+3}}
\sqrt{2(\nu^2+3)A^2+3(\nu^2-1)B} \,. \nonumber
\end{align}
Using these equations, the metric functions and their derivatives can  be expressed in terms of \(\mathcal{M}\) and \(\mathcal{J}\) as follows:
\begin{align}
N_\theta(r_+)\; &=\;\ell\Omega(\mathcal M,\mathcal J)
= \frac{2}{\nu\,C} \,,
\\
N_\theta'(r_+)
&= \frac{4}{\nu\,C^{2}}
\left[
1-\frac{1}{\nu^{2}C}\left(\frac{\nu^{2}+3}{2}\,A+\frac{3}{2}(\nu^{2}-1)\,C\right)
\right] \,,
\\
N_\theta''(r_+)
&= \frac{4\left( (\nu^2+3)A+3(\nu^2-1)C \right)^2-4\nu^2C\left( 2(\nu^2+3)A+9(\nu^2-1)C \right)}{\nu^5C^5} \,,
%\\
\end{align}
\begin{align}
R(r_+)
= \frac{\nu C}{2}\,,
\qquad \text{and}\qquad 
R'(r_+)
= \frac{(\nu^2+3)A+3(\nu^2-1)C}{4\nu C} \,.
\end{align}
In addition, the thermodynamic quantities—temperature \eqref{TMGwarpedte}, entropy \eqref{TMGwarpeden}, and angular velocity \eqref{TMGwarpedan}—can be expressed in terms of \(\mathcal{M}\) and \(\mathcal{J}\) as follows:
\begin{align}
T(\mathcal M,\mathcal J)
&= \frac{(\nu^{2}+3)\,\big(C-A\big)}{4\pi\,\nu\,\ell\,C}  \,,
\\
\mathcal S(\mathcal M,\mathcal J)
&= \frac{\pi\,\ell}{3\nu}\left[\,4\nu^{2}A + (5\nu^{2}+3)\big(C-A\big)\right] \,,
\\
\Omega(\mathcal M,\mathcal J)
&= \frac{2}{\ell\,\nu\,C}\,.
\end{align}
Substituting these expressions into Eq.~\eqref{tmgwarpedEOM} and multiplying $\rm{d} \mathcal{S}$ to both sides of the equation, we obtain the following field equation:
\begin{align}
    \mathcal{F}T\,\mathrm{d}\mathcal{S}
    &= \mathcal{F}\,\mathrm{d}\mathcal{M}
    - \frac{2\mathcal{F}}{\ell \nu C}
    \mathrm{d}\mathcal{J} \,,
\end{align}
where
\begin{align}
\mathcal F
&= \frac{20}{3}\,N_\theta'(r_+)\,R(r_+)^{2}\,R'(r_+) 
\;+\; 4\nu\,R'(r_+)
\;+\; \frac{4}{3}\,N_\theta''(r_+)\,R(r_+)^{3}  \,,
\\
&= \frac{\nu^2+3}{6\nu^2} \left[ 4\nu^2-(\nu^2+3)\left(1-\frac{A}{C} \right)^2 \right] \,.
\end{align}
Finally, we find that the above equation is equivalent to the first law of black hole thermodynamics:
\begin{align}
    T\,\mathrm{d}\mathcal{S}
    &= \,\mathrm{d}\mathcal{M}
    - \frac{2}{\ell\nu
    C}
    \mathrm{d}\mathcal{J} = \,\mathrm{d}\mathcal{M} - \Omega\,\mathrm{d}\mathcal{J}\,.
\end{align}

%\newpage

%\section*{References}
\bibliographystyle{unsrt}
\bibliography{ref}

\end{document}